\documentclass[nofootinbib]{revtex4}
\usepackage[english]{babel}
\usepackage{array,booktabs}   % http://ctan.org/pkg/{array,booktabs}
\usepackage{array} 
\usepackage{amsmath} 
\usepackage{relsize}
\usepackage{wrapfig}
\usepackage{calc}
\usepackage{pdflscape}
\usepackage{color}
\usepackage{float}
\usepackage[font=small,labelfont=bf]{caption}
\usepackage{graphicx}
\usepackage{amsmath}
\usepackage[nodisplayskipstretch]{setspace}

\usepackage{setspace}
\usepackage{tabularx}

\usepackage{float}
\usepackage{color}
\usepackage{amsmath}
\usepackage{float}
\usepackage{calc}
\usepackage{pdflscape}
\usepackage{color}
\usepackage{float}
\usepackage{graphics}
\usepackage{epsfig}
\usepackage{epstopdf}
\usepackage{amssymb}
\usepackage[font=small,labelfont=bf]{caption}
\usepackage{graphicx}
\usepackage{epstopdf}
\usepackage{appendix}
\usepackage{romannum}
\usepackage{soul}
\usepackage{color}

\definecolor{blizzardblue}{rgb}{0.67, 0.9, 0.93}
\definecolor{bubblegum}{rgb}{0.99, 0.76, 0.8}
\usepackage[urlcolor=blizzardblue]{hyperref}
\hypersetup{
    colorlinks=false,
    linkcolor=blue,
    filecolor=magenta,      
    urlcolor=bubblegum,
}
\begin{document} 
%\begin{frontmatter}

%\title{Nuclear Effects and Related Uncertainties at DUNE}
\title{Constraining nuclear effects in Argon using machine learning algorithms}
\author{Srishti Nagu$^{1}$ \footnote{E-mail: srishtinagu19@gmail.com}, Jaydip Singh$^{1}$ \footnote{E-mail: jaydip.singh@gmail.com}, Jyotsna Singh$^{1}$, R.B. Singh$^{1}$} 

\affiliation{Department Of Physics, University of Lucknow, Lucknow, India.$^{1}$}

\begin{abstract}
Neutrino oscillation experiments aim to measure the neutrino oscillation parameters with accuracy and achieve a complete understanding of neutrino physics. For determining the neutrino oscillation parameters, knowledge of neutrino energy is a prerequisite. But neutrino energy needs to be reconstructed, based on the particles in the final state that emerge out of the nucleus following a neutrino-nucleus interaction. Current and upcoming neutrino oscillation experiments use heavy nuclear targets (viz. Argon(Ar), Calcium(Ca), etc.) but the neutrino scattering with such targets becomes complicated as compared to that with a clean target like Hydrogen(H). This work explores the viability of using machine learning algorithms (MLA) in reconstructing neutrino energy. We use final state kinematics generated from two neutrino event generators viz. GENIE and GiBUU to train the MLA. We calculate the Ar/H ratio in an attempt to quantify nuclear effects in the Ar target. We observe a significant improvement in our results when we train the MLA by combining the FSI kinematics of neutrino interactions from both the neutrino event generators.
\end{abstract}
\maketitle

\section {Introduction}
Neutrino being intrinsically leptonic and neutral is the second most abundant particle in the Universe. It is a potential candidate for probing the complex hadronic structure as well as exploring physics beyond the Standard Model. Precision measurement of its properties and nature of interaction with the other nuclei is among the highest fundamental particle physics priorities. Several benchmark discoveries employing a variety of nuclear targets in the past decades have revealed several properties of the neutrino. Active neutrinos are known to exist in three flavors viz. electron neutrino ($\nu_{e}$), muon neutrino ($\nu_{\mu}$) and tau neutrino ($\nu_{\tau}$). A neutrino of a particular flavor, after traveling a certain distance, transforms into a neutrino of another flavor, this unique property is referred to as neutrino oscillation. A milestone was achieved with the discovery of neutrino oscillations that substantiated the tiny mass of neutrinos. The study of neutrino oscillation physics involves certain parameters i.e. mixing angles $\theta_{ij}$ where $j>i$=1,2,3 ($\theta_{12}, \theta_{13}, \theta_{23}$), Dirac phase $\delta_{CP}$ and mass squared differences i.e. $\Delta m_{31}^2$ known as atmospheric mass splitting and $\Delta m_{21}^2$ known as solar mass splitting. Some of the neutrino oscillation parameters have been measured precisely i.e. $\theta_{12}$, $\theta_{23}$, non-zero value of $\theta_{13}$ \cite{theta13_1,theta13_2,theta13_3} and magnitude of mass squared differences $\Delta m_{21}^2$, $|\Delta m_{31}^2|$. While the neutrino oscillation parameters which are yet to be discovered with precision are- (\romannum1) the value of dirac phase $\delta_{CP}$: which can lie in the range $-\pi < \delta_{CP} < \pi$ (\romannum2) the neutrino mass ordering i.e. sign of $\Delta m_{31}^2$: there is a possibility of neutrino mass ordering being 'normal' i.e. whether the neutrino mass order can be- $m_{1}\ll m_{2} \ll m_{3}$ known as normal mass hierarchy (NH) or the ordering can be 'inverted' i.e. the neutrino mass order can be- $m_{2}\approx m_{1} \gg m_{3}$ known as inverted mass hierarchy (IH), where $m_{i}$(i=1,2,3) are neutrino mass eigenstates (\romannum3) octant degeneracy i.e. whether the value of $\theta_{23}$ lies in the lower octant(LO) $0< \theta_{23} < \pi/4$ or higher octant(HO) $\pi/4 < \theta_{23} < \pi/2$. The current status of the neutrino oscillation parameters estimated by performing a global fit can be found in \cite{global}. Qualitative improvement in the measurement of the unknown neutrino oscillation parameters requires pinning down the systematic uncertainties which become more challenging with the use of heavy nuclear targets (mass number(A)$>$12). 

For precise determination of the neutrino oscillation parameters at neutrino oscillation experiments, it is crucial to accurately determine the energy of the incoming neutrino since the neutrino oscillation probability depends upon the neutrino energy spectrum. In beam-based collider/accelerator experiments we need to reconstruct the energy of the neutrino since neutrinos are produced as secondary decay products from primarily produced hadrons that lead to non-monoenergetic and broad neutrino beams. To reconstruct the energy of a neutrino, the energy of all the final state particles produced in a neutrino-nucleus interaction must be taken into account on an event-by-event basis. An obstacle for precise neutrino energy reconstruction from the final state particles is the inability to correctly identify the particle configuration, dynamics, and kinematics of a neutrino interaction within the nucleus. The particles that are produced at the initial neutrino nucleon interaction vertex within the nucleus are often modified inside the nuclear environment. When a neutrino interacts with a nucleon (which is not at rest) inside the nucleus certain nuclear effects \cite{nuceff} come into play such as fermi motion, Pauli blocking, multinucleon effects \cite{multinucleon}  at the initial interaction vertex. Moreover, as the particles produced at initial neutrino-nucleon interaction travel through the nucleus they re-interact with other nucleons and emerge out of the nucleus as different particles known as final state particles. Collectively, these effects are known as nuclear effects \cite{whitepaper}. As a result, the kinematics and dynamics of the real interaction remain unknown. Thus the physics of neutrino-nucleus scattering remains tangled and requires a good understanding of nuclear effects.

Neutrino owing to its weakly interacting nature and very low interaction cross-section is merely detectable. Thus to study neutrinos at neutrino experiments heavy nuclear targets are used which gives large statistics of neutrino interaction events resulting in reduced statistical uncertainty but at the same time leads to huge systematic uncertainties due to inherent nuclear effects. For extracting vital information from the large data samples collected from detectors at neutrino experiments, it is crucial to develop advanced reconstruction techniques. Machine learning algorithms(MLA) are an integral part of the analysis tools for the past two decades \cite{ml1}. Significant improvement has been witnessed in the physics reach of the analyses with the use of MLAs. Results obtained from previous generation experiments \cite{exp1,exp2,exp3,exp4,exp5} that involved the application of neural networks \cite{nn1,nn2} and Boosted Decision Trees(BDTs) led the use of MLAs at the Large Hadron Collider (LHC) \cite{lhc1}. MLAs also played a crucial role in the discovery of the Higgs Boson \cite{higgs1,higgs2}. For a much detailed review about the use of MLAs at LHC one can refer to \cite{lhc}. 

The upcoming and ambitious long-baseline neutrino oscillation experiment i.e. the Deep Underground Neutrino Experiment (DUNE) \cite{dune} is also using MLAs for data analysis. DUNE is currently in the process of being set up in the Fermilab, United States. It strives to achieve the most sought-after goals in the realm of particle physics i.e. the octant degeneracy, neutrino mass hierarchy, value of Dirac $\delta_{CP}$ phase. It's prime design consists of a Near Detector (ND) \cite{nd1,nd2,nd3,nd4} at Fermilab and a Far Detector (FD) \cite{dune} to be set up at Sanford Underground Research Facility (SURF), South Dakota. The design of ND is a fine-grained magnetic spectrometer while the FD will host four independent 10 kt each, Liquid Argon Time Projection Chambers (LArTPCs). Both the ND and FD will be employing Ar as a target with the aim of maximum cancellation of systematic uncertainties. A muon neutrino (or antineutrino) dominated beam will be focussed from Long-Baseline Neutrino Facility (LBNF) at Fermilab to both the ND and FD set up at a distance of 575 meters and 1300 km respectively. For data analysis at DUNE, advanced MLAs like Deep learning are being used for interaction vertex reconstruction \cite{dunemla1} and kinematic reconstruction \cite{dunemla2}. Deep learning is a type of machine learning method based on artificial neural networks where 'deep' refers to several layers in the network. Convolution Neural Network (CNN) \cite{cnn} is being used for classifying neutrino interactions \cite{dunemla3}. A CNN can be defined as a deep neural network that consists of several convolutional layers. The output from each layer acts as an input for the next layer. To classify neutrino interactions, this technique has proven to be better than the traditional event-based reconstruction methods \cite{refa,refb}. While reconstruction of neutrino energy and final state particle momenta with deep learning methods is still in progress.

To reconstruct the neutrino energy using MLAs, we have used a linear regression model and compared it with the calorimetric method of energy reconstruction. For our analysis we have selected two different simulation tools, GENIE (Generates Events for Neutrino Interaction Experiments) \cite{genie} (version-2.12.6) and GiBUU (Giessen Boltzmann-Uehling-Uhlenbeck) \cite{gibuu} (version-2019). Both GENIE and GiBUU are based on different physics approach i.e. they employ different nuclear models to incorporate nuclear effects and various neutrino interaction processes. In an attempt to constrain systematic uncertainties in Ar, we have calculated the ratio of Ar/H using calorimetric and MLA. Further, we have taken a step to check the multi-generator approach and combined the final state interaction (FSI) kinematics as obtained from both generators. Our results show that on adopting a multi-generator approach systematic uncertainties were significantly reduced.

The paper is organized in the following sections: We describe the neutrino event generators GENIE and GiBUU used in this work in Section II. We outline the simulation, experimental details, and the analysis strategy in Section III. The results are discussed in Section IV followed by the Conclusions in Section V.

\section{Neutrino event simulation tools}
A neutrino event generator simulates neutrino-nucleus interactions for a particular detector setting and propagates the products out of the struck nucleus. These generators take into account, both the initial neutrino-nucleon interaction and the final state interactions. Event generators must be efficient enough to simulate interactions on a wide range of nuclei, for a broad neutrino energy spectrum, for all the neutrino (anti-neutrino) flavors, and must have accurate energy-dependent cross-section. The role of the event generator becomes crucial as it provides information about the physics of the interaction taking place. But we lack complete information about the initial neutrino-nucleus interaction and the subsequent re-interactions taking place within the nucleus that lead to modified outgoing products. Thus different neutrino event generators incorporate different nuclear and physics models to describe the nucleus and the various energy-dependent interaction processes. Different neutrino experiments use different neutrino event generators, for eg. NEUT \cite{neut} is used by the T2K experiment, GENIE \cite{genie} is used by Fermilab experiments like NOvA \cite{7e}, MicroBooNE \cite{7d}, MINERvA \cite{7b}, T2K\cite{7f} and MINOS\cite{7c}. NuWRO \cite{nuwro} is being used for the comparison of experiments and calculations. Another generator named, GiBUU \cite{gibuu} can be used for a wide range of interactions between a nucleus- and -neutrino/electron/photon/pion/nucleus. 

Our selected event generators i.e. GENIE and GiBUU, considerably differ in the selection of nuclear models and description of above mentioned neutrino-nucleus interactions. GENIE is developed in C++ language and is a ROOT\cite{7a} based neutrino event generator designed using object-oriented methodologies. Whereas, GiBUU, written in FORTRAN routines, is based on a coupled set of semiclassical kinetic equations. These equations describe the dynamics of a hadronic system in phase space and time. 

Some differences and similarities between these two event generators can be outlined as follows. For nuclear density distribution both the generators use Woods-Saxon parametrization\cite{woodsaxon}. The value of axial mass used by GiBUU is $M_{A}$= 1 $GeV/c^{2}$ while the value of axial mass in GENIE is 0.99 GeV/$c^{2}$. The vector form factor, BBBA07 \cite{refGibuuVct} is used by GiBUU while GENIE uses BBBA05 \cite{refGenieVct}. To describe the nuclear density distribution, both the generators use Woods-Saxon parametrization \cite{woodsaxon}. For simulating QE interactions that are dominant below 1 GeV, RFG model\cite{refrfg} is used to describe the nuclear structure by both the generators. But the RFG model used in GiBUU has an additional density-dependent mean-field potential term in which all nucleons are assumed to be bound while in GENIE it is based on the model suggested by A. Bodek and J.L. Ritchie \cite{ABodak} that includes short-range nucleon-nucleon correlations. The QE scattering in GENIE is modeled according to the Llewellyn Smith model\cite{qemodel}. For the description of QE cross-section in GiBUU, one can refer to \cite{minibooneQE,moselQE}. Below 1 GeV, there is an interplay of other interactions also. Nucleon-nucleon correlations involving two-particle two-hole (MEC/2p2h) excitations mainly populate in this low energy region \cite{2p2h}. 

The resonance interaction dominates in the energy range from 0.5 $< E_{\nu} <$ 10 GeV. There are 13 variants of resonance modes in GiBUU and it is the MAID analysis \cite{refGibuuMAID1,refGibuuMAID2} of the electron scattering data that defines vector form factors for all the modes. GENIE accommodates 16 resonance modes defined by Rein Sehgal model\cite{RShegal}. At higher energies, DIS interaction dominates where the interaction is at the quark level. GiBUU and GENIE implement PYTHIA \cite{gibuudis} and Bodek and Yang \cite{bodekyang} respectively to describe the DIS process. GENIE incorporates Aivazis, Olness, and Tung model \cite{discharm} also, particularly for DIS charm production.

The initially produced particles at the neutrino-nucleon interaction vertex are further modified due to re-interactions with other nucleons within the nucleus, such re-interactions are referred to as Final State Interactions (FSI). Both GENIE and GiBUU apply different approaches to model FSI. FSI in GENIE is modeled using the simulation package INTRANUKE \cite{int1,int2,dytman1,dytman2}. GiBUU models the FSI by the semi-classical BUU equations. A mean-field potential and collision terms keep different particle species coupled to each other. Further information on modeling differences, corrections, and improvements being implemented in the two generators, one can refer to \cite{ref45,refGenieMEC,refGibuu2p2h}.

\section{SIMULATION AND EXPERIMENTAL DETAILS}
For simulating a DUNE-like experiment, we have considered the DUNE-ND flux \cite{ndflux} that spans over an energy range of 0.125-10 GeV as shown in Figure 1. In this energy range, there is an interplay of several interaction processes viz. quasi-elastic (QE), resonance (RES) from $\Delta$ resonant decay, and contribution from higher resonances, two particle-two hole(2p2h/MEC), and deep inelastic scattering(DIS) interaction processes. We have considered all these interaction processes in our analysis. 

The incoming neutrino beam is dominated by muon flavor which is generated at the LBNF facility. At LBNF, a beam of protons in the energy range 60-120 GeV is produced at the main injector accelerator and is made to smash a graphite target. This results in the production of pions and kaons which are directed to a decaying pipe with the help of magnetic horns. These pions and kaons further decay into neutrinos and leptons of all flavors. Finally an intense, wide-band, $\nu(\bar\nu)$ beam is produced with an initial beam power of 1.2 MW, producing 1.1 $\times$ $10^{21}$ protons on target per year \cite{dune}. 

\begin{figure}
\centering\includegraphics[scale=.44]{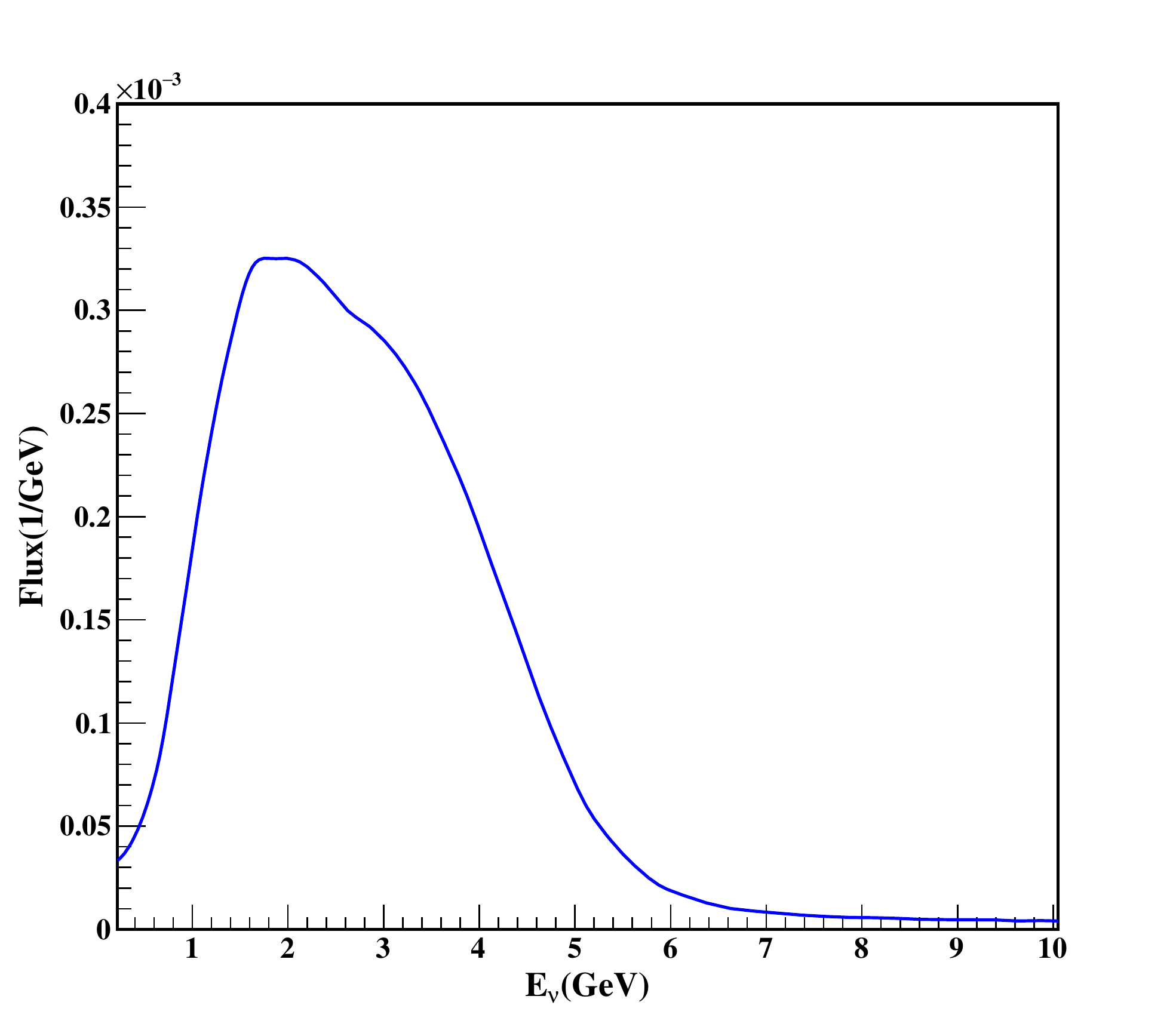}
%\centering\includegraphics[scale=.44]{/home/srishti/JDPapers/ErecMachLearnPaper/fdflux.pdf}
\caption{The DUNE-ND flux as a function of neutrino energy used in our work.}
\end{figure}

Inclusive event samples of 1 million $\nu_{\mu}$-Ar and $\nu_{\mu}$-H each, are generated using both the neutrino event generators i.e. GENIE and GiBUU. The choice of nuclear targets is such so that we can quantify nuclear effects in Ar. As we know DUNE-ND and FD will employ liquid Ar in both the detectors, it is crucial to study neutrino-Ar scattering. We have selected H because it is a single proton and hence it is devoid of nuclear effects. It is very important to have a control sample that can directly constrain nuclear effects which can help in reducing systematic uncertainties. Comparing measurements on Ar and H will impose stringent constraints on the nuclear smearing in Ar. The Straw Tube Tracker (STT), a component in SAND \cite{nd4}, which is one of the components in the DUNE-ND will be studying $\nu$-Ar and $\nu$-H interactions along with other nuclear targets (For eg. Ca(A=40), C(A=12)) as well. STT will be using H in the form of 'solid hydrogen' and will obtain interaction kinematics with H by subtracting interactions on C$H_{2}$ \cite{roberto} and Carbon. This is the motivation behind our selection of targets as, without a complementary H target, it is difficult to achieve precision measurement free from nuclear smearing. Although the DUNE collaboration does not have any plan to use an H target in any form in the FD module as planned for the ND complex (SAND), but for simulation purposes, we have replaced Ar with an H target. We have estimated the probability with H target as a function of reconstructed neutrino energy using a single generator and multi-generator approach as well. To quantify the amount of nuclear effect we have further used it for calculating the Ar/H ratio as presented in Figures 6 and 7.

\subsection{Analysis strategy}
The following section describes the two analysis techniques that we have applied in our work i.e. Calorimetric and Linear regression methods. 
\subsubsection{Calorimetric technique}
The calorimetric method \cite{c1,c2} of neutrino energy reconstruction estimates the neutrino energy by adding the energies deposited by all the particles observed in the detector. This method in comparison to the traditional kinematic method \cite{k1,k2,k3,k4} of neutrino energy reconstruction is more effective for mainly two reasons. Firstly, the kinematic method is only restricted to the QE interaction process as it is based on lepton kinematics (outgoing muon) only while the calorimetric method applies to all types of interaction processes. Secondly, the calorimetric method is in a way insensitive to nuclear effects while the kinematic method is more prone to nuclear effects for interaction processes other than QE. But only in an ideal detector where there will be no missing energy i.e. even the particles with the lowest energies are detected, then only the calorimetric method will reconstruct effectively. Although the impact of FSI will still be prevalent in neutrino interactions due to which the energy transferred to the nucleus may get redistributed to several particles but the total energy will remain the same \cite{mosel}. 

For calculating the neutrino energy ($E_{\nu}^{Cal}$) using the calorimetric approach, $E_{\nu}^{Cal}$ \cite{c1} can be calculated as:
\begin{equation}
E_{\nu}^{Cal} = E_{l} + \Sigma_{i} T_{i}^{n} + \epsilon_{n} + \Sigma_{m} E_{m}
\end{equation}

where $E_{l}$ is the energy of the outgoing final state lepton, $T_{i}^{n}$ is the kinetic energy of the outgoing nucleons (protons and/or neutrons), $\epsilon_{n}$ represents the corresponding separation energies of the outgoing nucleons and $E_{m}$ is the total energy of any other particle produced. The calorimetric energy in our work is calculated using Eq.(1).

\subsubsection{Multivariate linear regression}
We have applied the multivariate linear regression model which comes under supervised learning. In supervised learning, we have the required dataset that is the input and have an idea about the output. Linear regression predicts a real-valued output based on an input value. Our work gives multiple values, known as feature variables in the input, which is termed as multivariate analysis. A python-based script is developed that takes the FSI kinematics (features for our algorithm) as an input to train the algorithm and predict the neutrino energy. The multiple features are the number of particles per event, momenta of all the outgoing particles per event, true neutrino energy, the energy of outgoing muon ($E_{\mu}$), energy transfer ($q_{0}$), visible momenta of each particle, squared four momenta ($Q^{2}$), and reconstructed neutrino energy. This dataset is further divided into a training set and testing set as 70$\%$ and 30$\%$ respectively. The training set is used to train the MLA and the testing set is used to check the predictions done by the MLA. The algorithm learns from the input variables about the reconstructed neutrino energy per event and then predicts the reconstructed neutrino energy in the output based on its learning from the feature variables provided to it. The regression algorithm performs a mapping between a dependent and an independent variable which is a continuous mapping and can be used to predict the energy of a particle. 

%event,mu,pip,pim,pi0,proton,neutron,photon,Pmu,Ppip,Ppim,Ppi0,Ppro,Pneutr,Enu,El,q0,Pvis,Q2,q3,Erec

\begin{figure}
\centering\includegraphics[scale=.44]{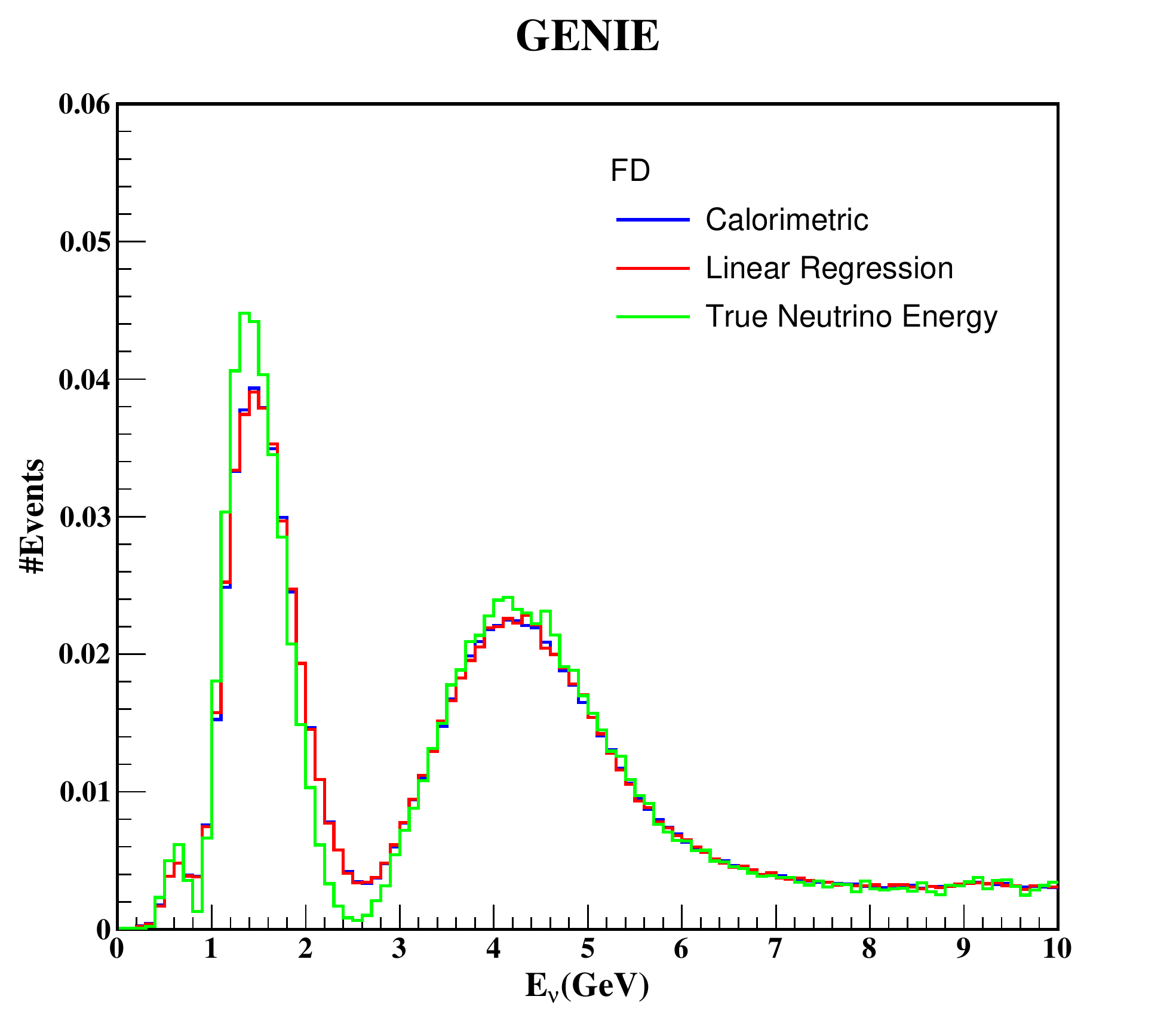}
\centering\includegraphics[scale=.44]{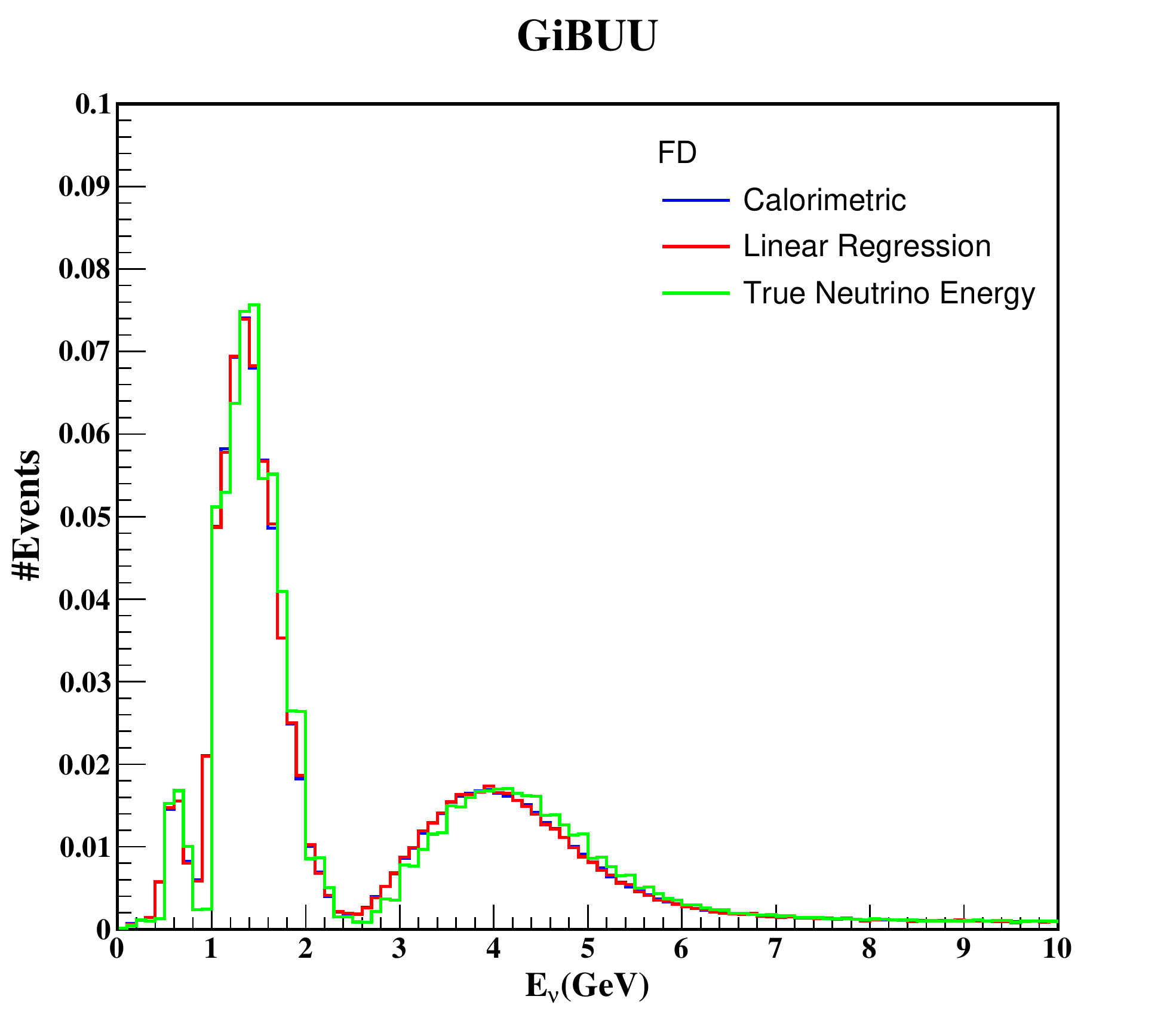}
\caption{Event distribution at the FD as a function of reconstructed neutrino energy represented by blue (Calorimetric method), red (linear regression method) lines, and as a function of true neutrino energy represented by green lines for Ar target from GENIE and GiBUU in the left and right panels respectively.}
\end{figure}

\begin{figure}
\centering\includegraphics[scale=.44]{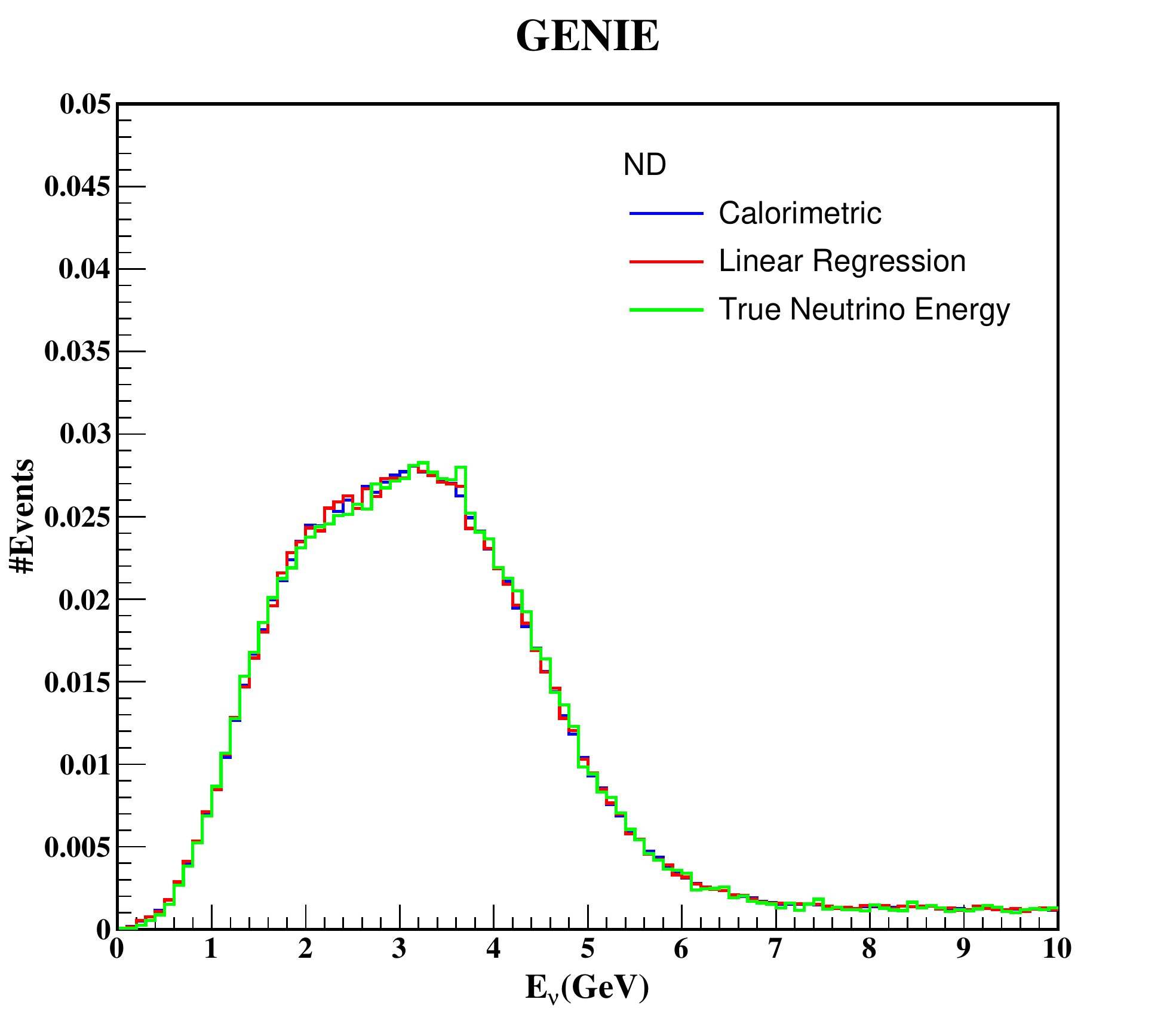}
\centering\includegraphics[scale=.44]{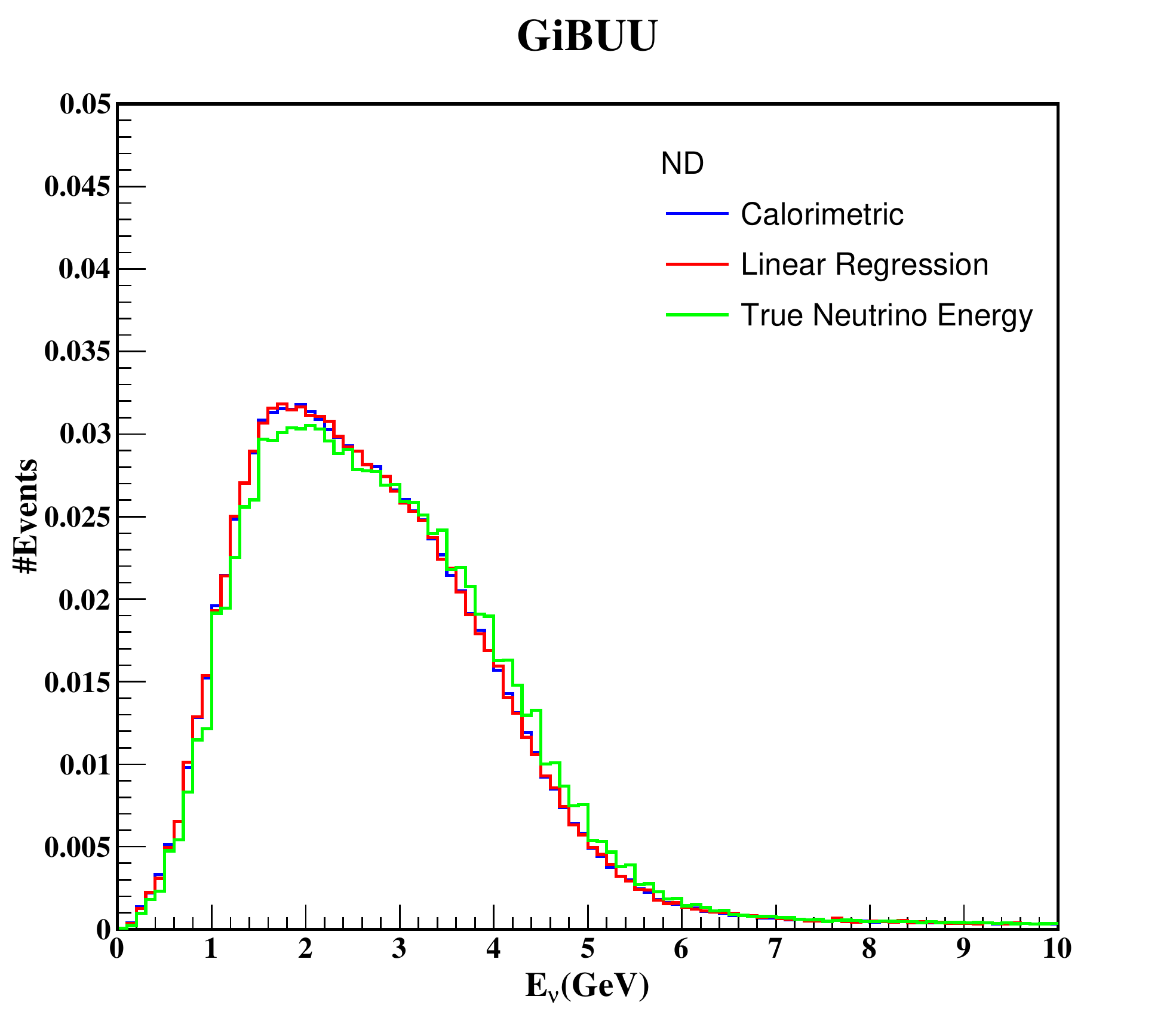}
\caption{Event distribution at the ND as a function of reconstructed neutrino energy represented by blue (Calorimetric method) and red lines respectively and as a function of true neutrino energy represented by green lines for Ar target from GENIE and GiBUU in the left and right panels respectively.}
\end{figure}

\begin{figure}
\centering\includegraphics[scale=.44]{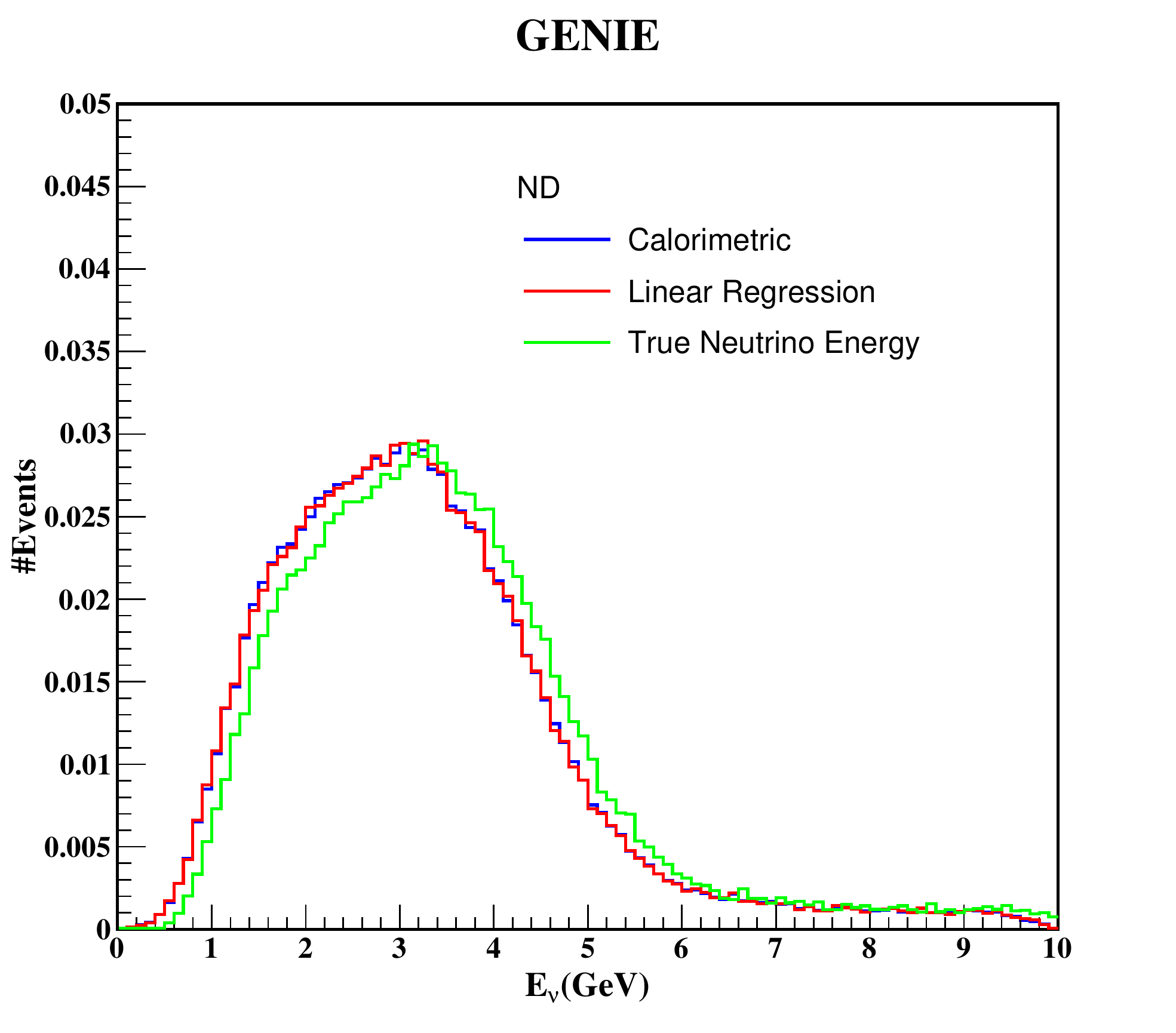}
\centering\includegraphics[scale=.44]{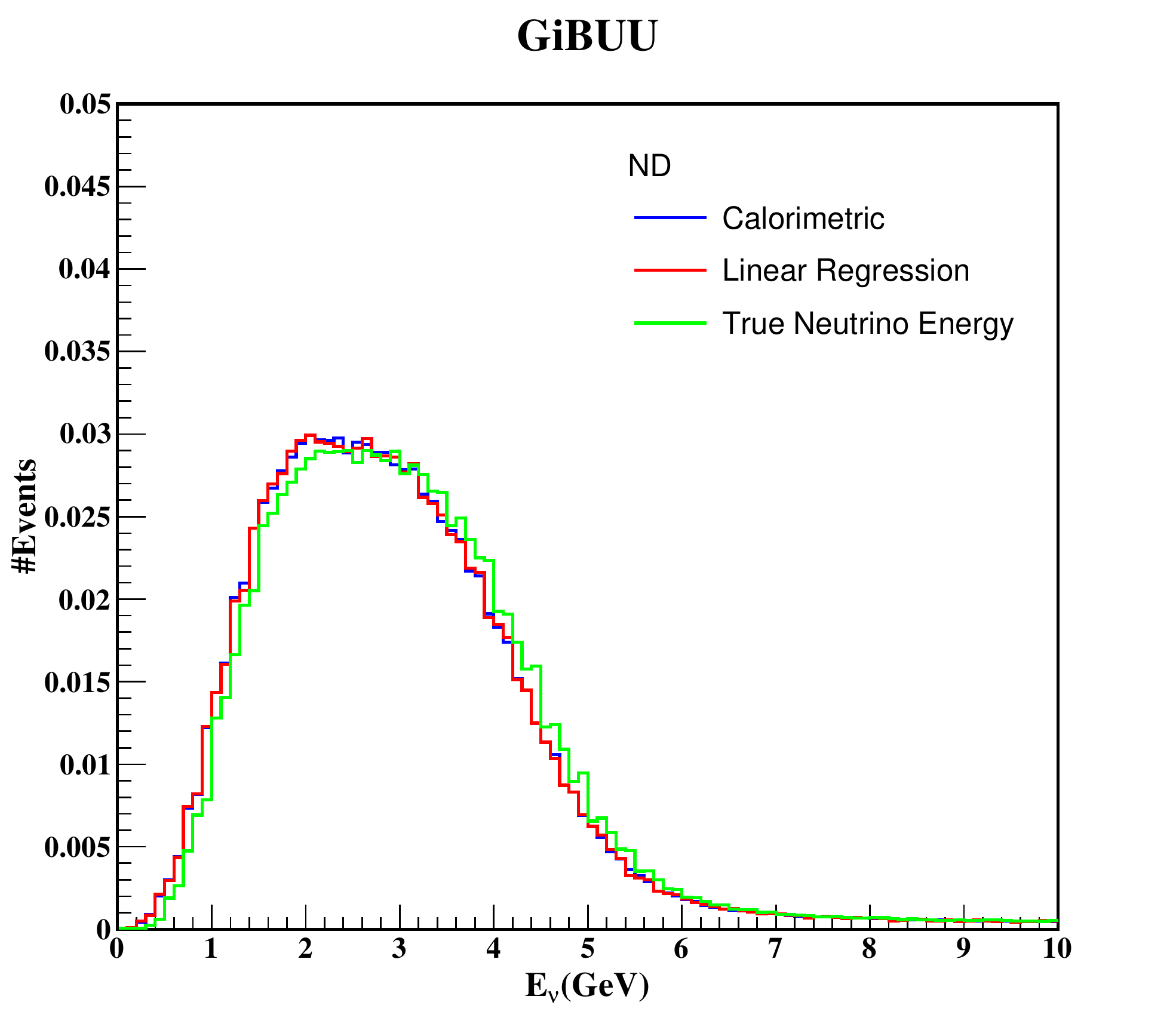}
\caption{Event distribution at the ND as a function of reconstructed neutrino energy represented by blue (Calorimetric method) and red (linear regression method) lines respectively and as a function of true neutrino energy represented by green lines for H target from GENIE and GiBUU in the left and right panels respectively.}
\end{figure}

\begin{figure}
\centering\includegraphics[scale=.44]{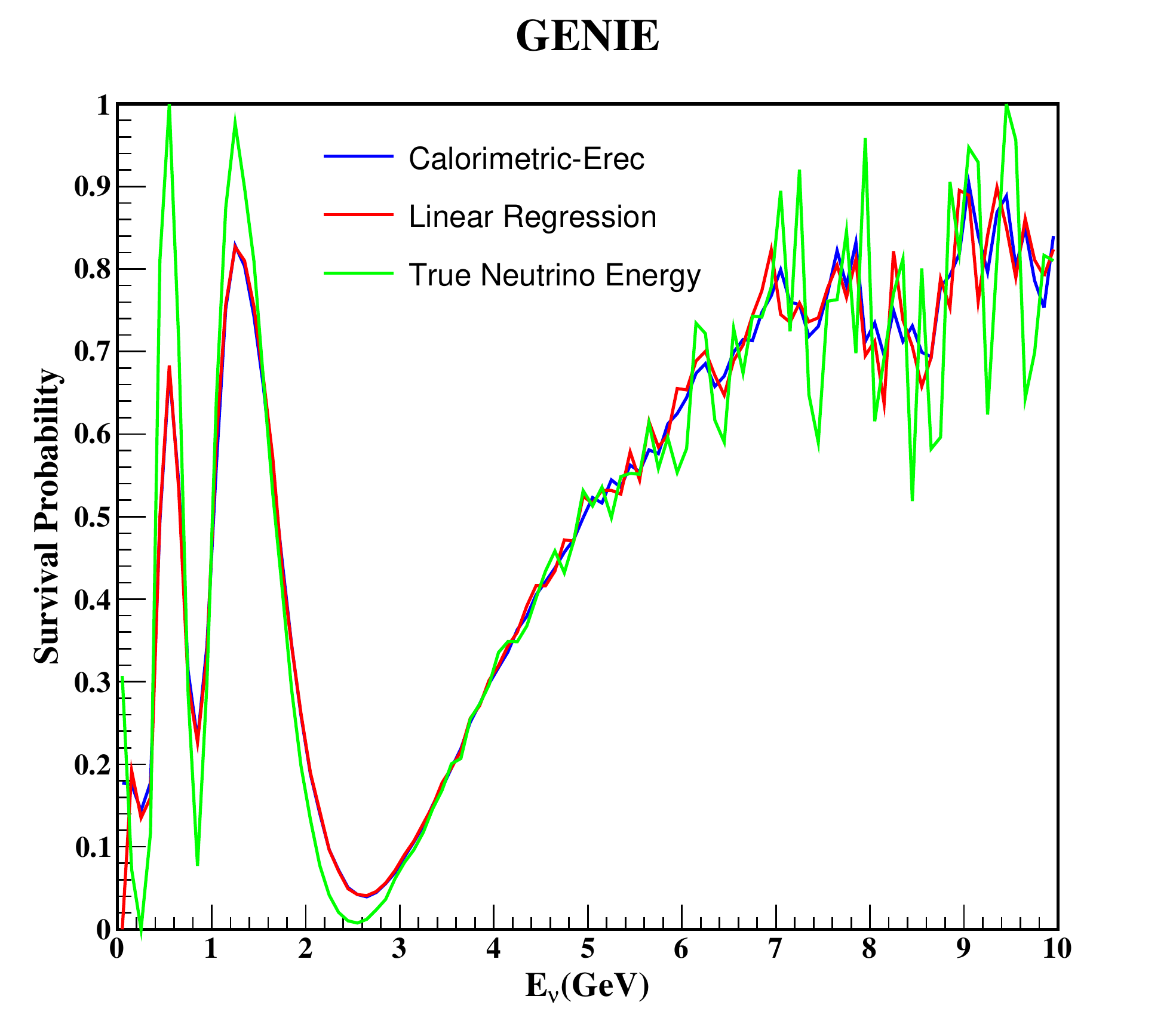}
\centering\includegraphics[scale=.44]{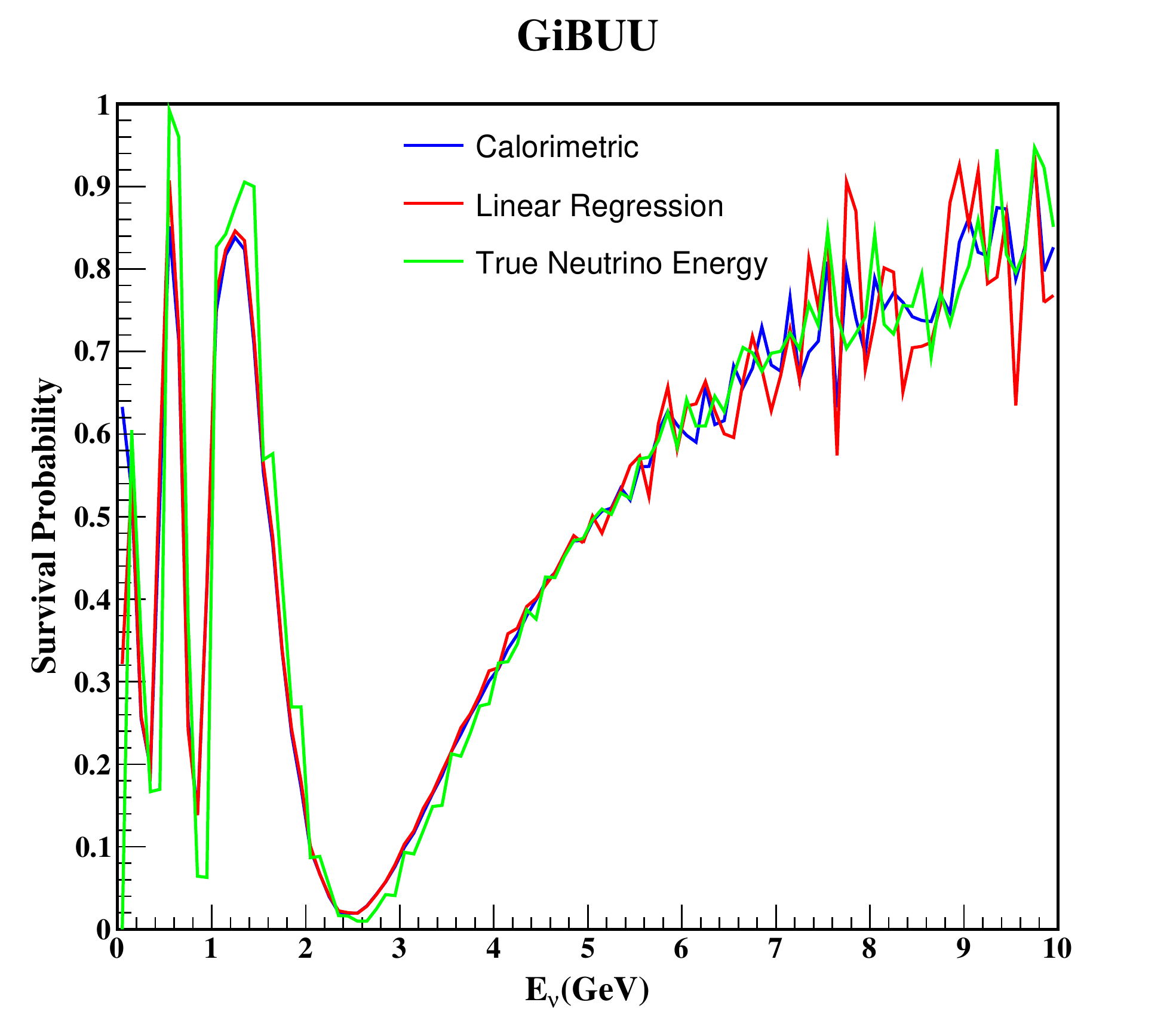}
\caption{Muon survival probability as a function of reconstructed neutrino energy represented by blue (Calorimetric method) and red (linear regression method) lines respectively and as a function of true neutrino energy represented by green line for Ar target from GENIE and GiBUU in the left and right panels respectively.}
\end{figure}

\section{Result and discussion}
On testing the viability of the linear regression method for reconstructing the neutrino energy, we find that the method can competently reconstruct neutrino energy as good as the calorimetric method. Figures 2 and 3, represent the $\nu_{\mu}$-Ar event distribution as a function of reconstructed neutrino energy at the FD and ND respectively. Both the figures display the event distribution calculated using the calorimetric method (blue lines) and linear regression method (red lines) by GENIE and GiBUU in the left and right panels respectively. The green line in both figures represents event distribution as a function of true neutrino energy.  Figure 4 represents the $\nu_{\mu}$-H event distribution as a function of reconstructed neutrino energy at the ND only. We again notice that the reconstructed neutrino energy is very well reconstructed using the linear regression method. 

Figure 5 represents the muon survival probability by taking the ratio of FD and ND event distributions as obtained from GENIE and GiBUU, in the left and right panels respectively for Ar. The blue line represents the muon survival probability as calculated from the calorimetric method and the red line represents the same as calculated using the linear regression method. We notice that both the blue and red lines completely overlap each other indicating that the linear regression method works well for neutrino energy reconstruction. 

In an attempt to estimate the impact of systematic uncertainties arising due to nuclear effects in neutrino energy reconstruction, we have taken the ratio of muon survival probabilities for Ar over H and presented it in Figures 6 and 7, for GENIE and GiBUU respectively. From Figure 6, we note that the ratio from GENIE as represented in the left panel has a lot of fluctuation up to 3 GeV as calculated from both the calorimetric (blue circles) and linear regression (red circles) methods. We also note two sharp peaks, one below 1 GeV and another around 2 GeV. Ratio when calculated as a function of true neutrino energy (represented by green circles) shows lesser fluctuation as compared to the calculation performed with reconstructed neutrino energy. On the other hand, the right panel of Figure 6, which represents the ratio calculations from GiBUU shows lesser fluctuations up to 3 GeV, and no sharp peaks are observed as in the case with GENIE.  
\begin{figure}
\centering\includegraphics[scale=.44]{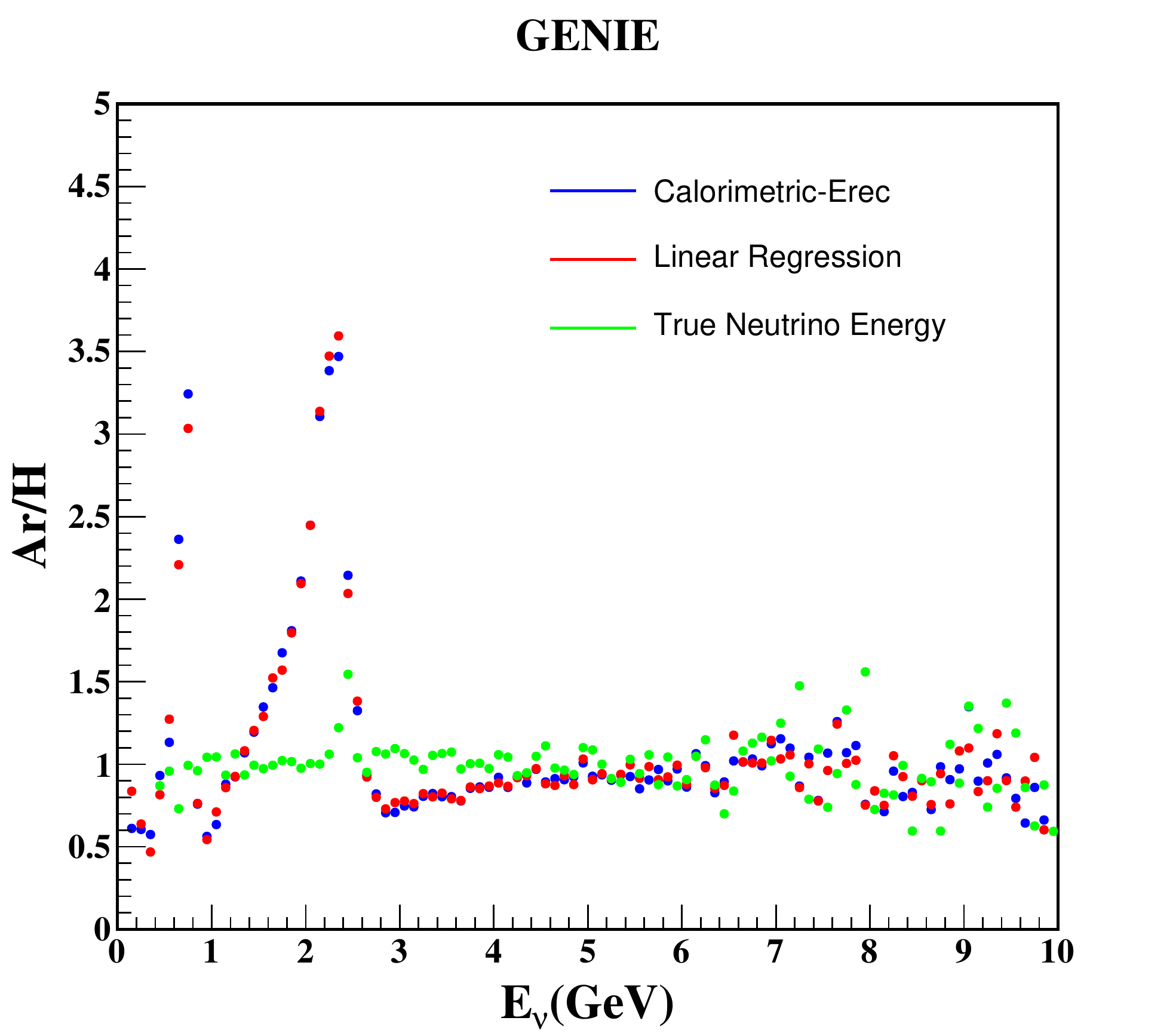}
\centering\includegraphics[scale=.44]{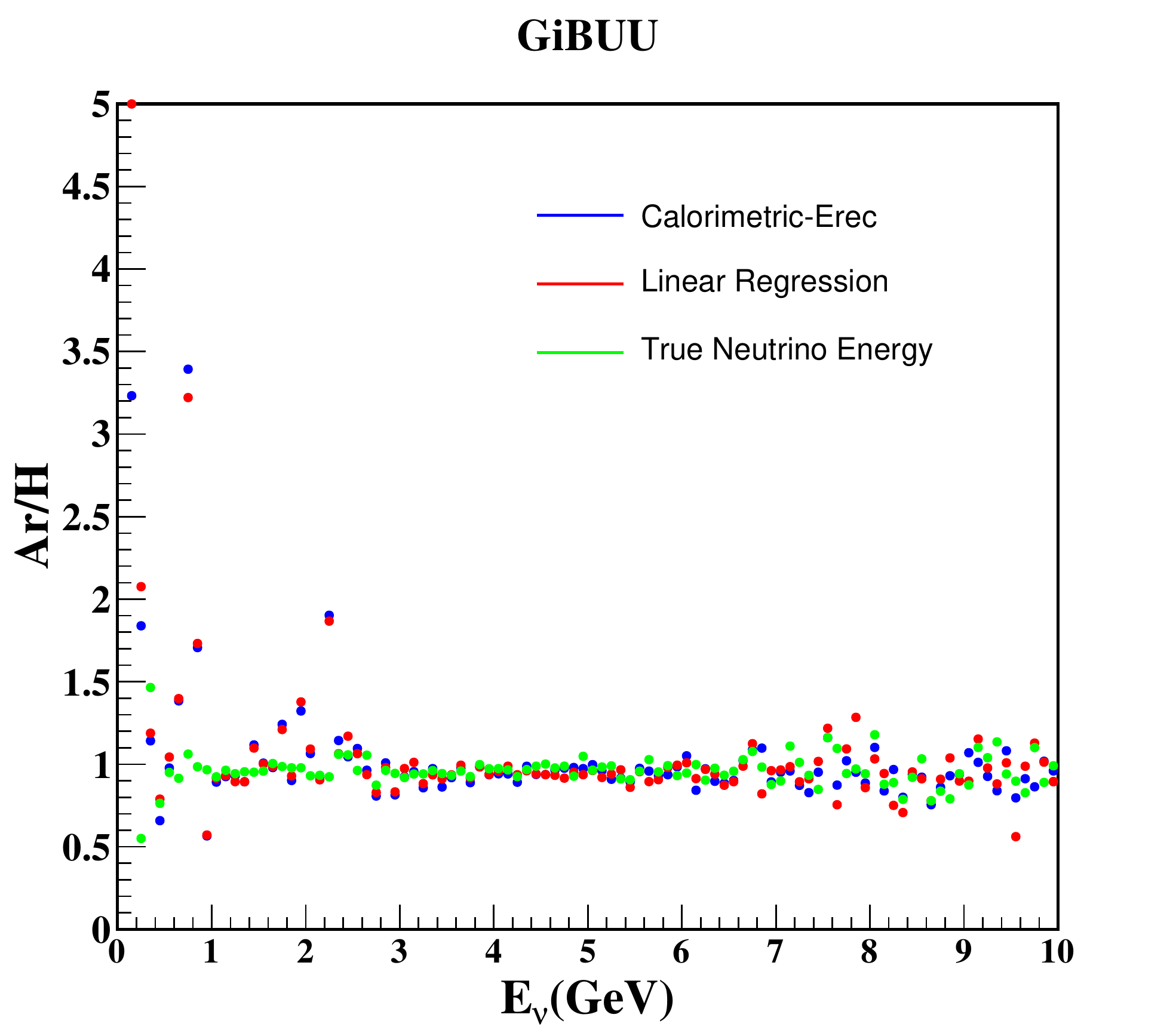}
\caption{Ar/H Ratio of muon survival probability by Calorimetric method (blue line) and linear regression method (red line) from GENIE and GiBUU in the left and right panels respectively.}
\end{figure}

\begin{figure}
\centering\includegraphics[scale=.44]{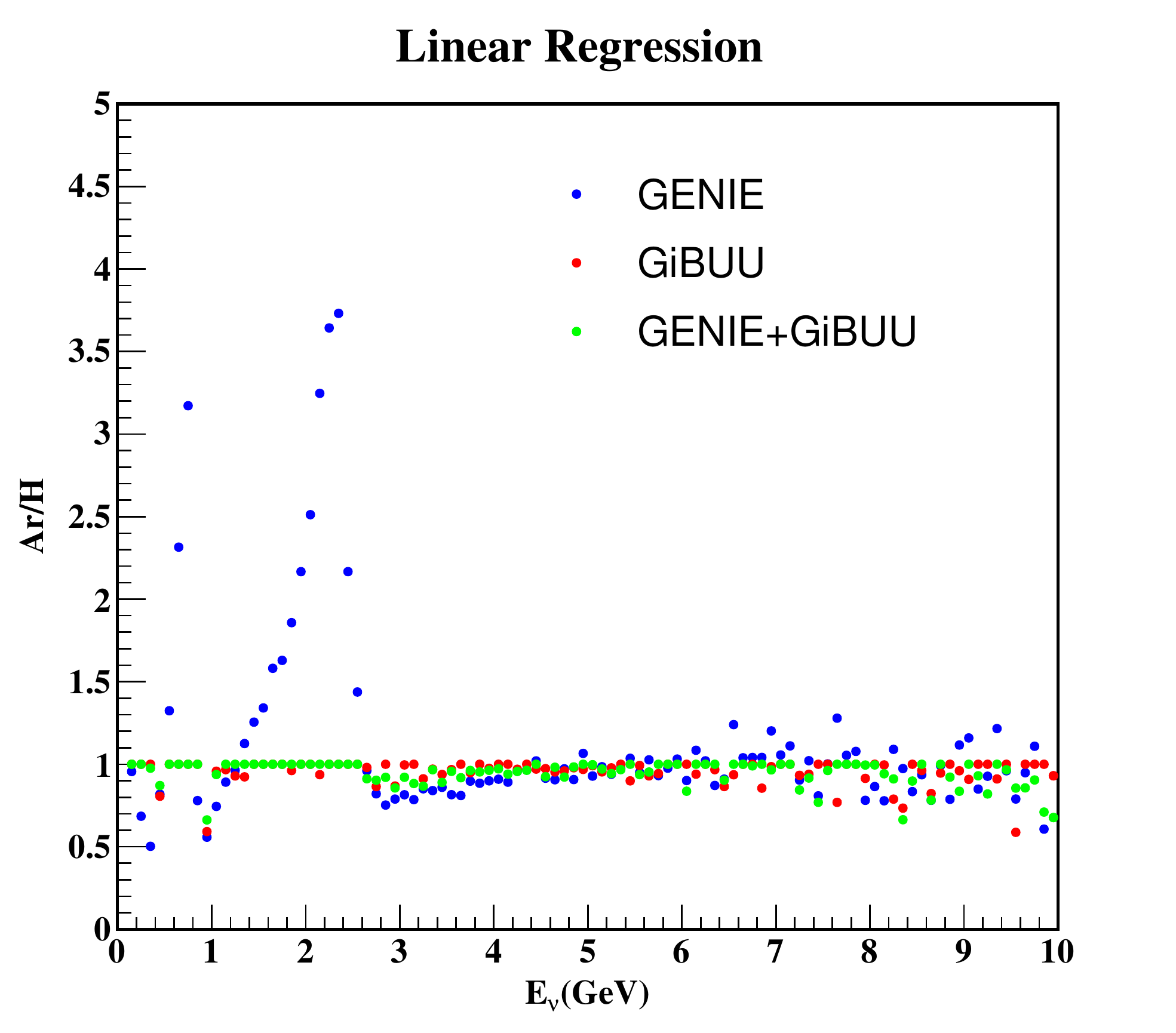}
\caption{Ar/H ratio from GENIE (blue circles), GiBUU (red circles) and GENIE+GiBUU (green circles) using only linear regression.}
\end{figure}

In Figure 7, we have presented three ratios i.e. from GENIE (blue circles), GiBUU (red circles), and GENIE+GiBUU (green circles) using only the linear regression model. For calculating the Ar/H ratio for the GENIE+GiBUU dataset, we simply combined the final state kinematics as obtained from the two generators and performed all the calculations again. From Figure 7, we can observe that the GENIE+GiBUU dataset shows a better result as the ratio is around unity for the entire energy range.

\section{Conclusion}
Neutrino interaction event rates are a convolution of neutrino flux, neutrino-nucleon cross-section, nuclear
effects and detector response. The first three factors depend upon the incoming neutrino energy whereas the detector response depends upon the efficiency and threshold of the detector. Experimentalists use event generators to connect the observed topologies and interaction kinematics. Due to scarcity of data, there are cross-sectional uncertainties where unknown fundamental neutrino scattering kinematics further distort vital information that lead to systematic uncertainties. To construct a precise nuclear model, we need exact information about all the convoluted factors. 

We find from Figure 6, that the two widely used neutrino event generators, GENIE and GiBUU have non-negligible systematic uncertainties in their nuclear models and especially in the energy range from 1-3 GeV. QE and RES interaction processes are the most dominant interactions in this energy range. This energy range has other interaction processes like multinucleon, 2p-2h, and coherent interactions that are still not modeled properly. We can note that GENIE results have more fluctuations than GiBUU results, which indicates a larger amount of uncertainties in the GENIE in the low energy range. From Figure 7, we observe a significant improvement in the Ar/H ratio, represented by green circles. This shows that on combining the FSI kinematics from both the generators, fluctuation in the ratio was minimal, indicating a significant reduction in the systematic uncertainties. A study of high-statistics $\nu$-H scattering data is necessary to supplement our knowledge of the poorly known nuclear effects. Through this work, we may conclude that a multi-generator and multi-target approach might be useful to constrain nuclear effects in the future long-baseline neutrino experiments like DUNE.

\section{Acknowledgment} 
This work is supported by the Department of Physics, University of Lucknow, Lucknow.

\end{document}